# Simplified radar architecture based on information metasurface

Si Ran Wang[1], Zhan Ye Chen[1,2,3,*], Shao Nan Chen[1,2,3], Jun Yan Dai[1,2,3], Jun Wei Zhang[1,2,3], Zhen Jie Qi[1,2,3], Li Jie Wu[1,2,3], Meng Ke Sun[1,2,3], Qun Yan Zhou[1,2,3], Hui Dong Li[1,2,3], Zhang Jie Luo[1,2,3,*], Qiang Cheng[1,2,3,*], Tie Jun Cui[1,2,3,4,*]

1. State Key Laboratory of Millimeter Waves, Southeast University, Nanjing 210096, China
2. Institute of Electromagnetic Space, Southeast University, Nanjing 210096, China
3. Frontiers Science Center for Mobile Information Communication and Security, Southeast University, Nanjing 210096, China
4. Suzhou Laboratory, Suzhou 215164, China

These authors contributed equally: *Si Ran Wang, Zhan Ye Chen, Shao Nan Chen*

E-mail: chenzhanye@seu.edu.cn; zjluogood@seu.edu.cn; qiangcheng@seu.edu.cn; tjcui@seu.edu.cn

## Abstract

Modern radar typically employs a chain architecture that consists of radio-frequency (RF) and intermediate frequency (IF) units, baseband digital signal processor, and information display. However, this architecture often results in high costs, significant hardware demands, and integration challenges. Here we propose a simplified radar architecture based on space-time-coding (STC) information metasurfaces. With their powerful capabilities to generate multiple harmonic frequencies and customize their phases, the STC metasurfaces play a key role in chirp signal generation, transmission, and echo reception. Remarkably, the receiving STC metasurface can implement dechirp processing directly on the RF level and realize the digital information outputs, which are beneficial to lower the hardware requirement at the receiving end while potentially shortening the time needed for conventional digital processing. As a proof of concept, the proposed metasurface radar is tested in a series of experiments for target detection and range/speed measurement, yielding results comparable to those obtained by conventional methods. This study provides valuable inspiration for a new radar system paradigm to combine the RF front ends and signal processors on the information metasurface platform that offers essential functionalities while significantly reducing the system complexity and cost.

## Introduction

Radar stands for Radio Detection and Ranging, and is one of the most important technologies used for detecting and measuring the distance and speed of targets[1-3]. Since its invention during World War II, radar has made significant progress along with the advancement of technologies such as computers and materials. Today, radar is extensively used across diverse industries, encompassing military, civil, and commercial areas, evolving towards collaboration, intelligence, integration, multi-functionality, streamlining, and cost-effectiveness.

Conventionally, modern radar systems are made in chain architectures that comprise radio-frequency (RF) transmitters and receivers, and digital signal processors[2,4]. The radar signal generation relies on voltage-controlled oscillator (VCO) or direct digital synthesis (DDS)[5,6] technologies by using active RF components such as mixers and amplifiers, which are associated with high cost, high power consumption, high complexity, and small flexibility. In addition, the separation between the antennas and the RF front end reduces the integration degree of systems. At the receiver, radar echoes are captured by antennas and subsequently converted from analog to digital formats through analog-to-digital (AD) converters[1,3]. This conversion process is critically reliant on the performance of the sampling components, which can become a limiting factor, particularly in broadband radar systems. The transition from RF reception to digital processing not only introduces additional time delays but also increases the cost, complexity, and power consumption of the system. Furthermore, this transition can introduce noise, which may affect the quality of the signals. Therefore, it is of great significance to investigate high-efficiency systems with low cost, simplified architecture, and high flexibility.

The rapid progress in metasurface science and technologies[7,8] opens up opportunities for new radar designs. As a two-dimensional (2D) version of metamaterial[9-14], the metasurface consists of sub-wavelength elements arranged periodically or quasi-periodically. Over the past decade, the metasurface has demonstrated strong abilities in tailoring electromagnetic (EM) waves, involving beam steering[15,16], polarization conversion[17,18], nonreciprocal transmission[19-22], vortex-wave generation[23,24], and absorption[25,26], to name a few. With the integration of electrically switching and tunable components and field programmable gate array (FPGA), a new kind of metasurfaces, digital coding and programmable metasurfaces, was proposed[9], making it highly flexible in manipulating the EM waves in the time, frequency, and spatial domains in real time by properly designing the time-coding, space-coding, and space-time-coding (STC) sequences[9,28,30]. The programmable metasurface was further extended to information metasurface due to its special capability to control the EM waves and modulate digital signals simultaneously[27,28]. The programmable and information metasurfaces have been extensively investigated in many scenarios, such as harmonic beamforming[30], high-efficiency nonlinear frequency conversions[31] 错误!未找到引用源。, and EM imaging[33-34]. In the field of wireless communication, the programmable metasurfaces have been attracting significant attention due to their abilities to reshape the channel environments[35,36]. Moreover, with the introduction of time modulation strategies, the information metasurfaces can directly modulate the digital information on the harmonics, eliminating the need for mixers and amplifiers commonly used in the traditional transmitters[37-41].

However, deep integration of metasurfaces in radar systems is rarely reported. Low-cost phase-controllable metasurfaces were developed to replace the conventional phased arrays[42,43] and frequency diverse arrays (FDAs)[44-46]. Recently, benefiting from the abilities to produce nonlinear harmonics, the information metasurfaces were employed to generate frequency-modulated continuous wave (FMCW)[47] and micro-Doppler signals[48], but these works were only focused on the transmitting end. Till now, very little attention is given to the use of metasurfaces at the receiving end. This is primarily due to two factors. Firstly, conventional radar receivers primarily perform signal processing in the digital domain, which do not take full advantage of real-time RF processing potentials offered by the information metasurfaces. Secondly, dechirp operation requires the metasurfaces to possess full-phase control capability, but few metasurfaces are capable of fulfilling this requirement. Recent research has explored the potential to use metamaterial or metasurface for computing, including various operations such as differentiation, integration, convolution, discrete Fourier transform (DFT), and matrix equation solving[49-54]. However, these studies required additional RF circuit components[49,50] and complex iterative algorithms[51,52] or artificially intelligent networks[53,54]. To the best of our knowledge, there has been no advancement in the use of metasurface for radar receivers. Hence, the research on metasurface-based radar systems remains stagnant.

In this work, we propose a simplified radar architecture that leverages STC information metasurfaces as the key components in the chain of signal generation, transmission, reception, and processing. At the transmitting end, the STC metasurface performs not only beamforming but also radar-signal generation, which holds the following two main advantages. Firstly, it offers a versatile approach to signal and parameter regulation through digitally controlling signals. Secondly, without using the VCO or DDS technologies that operate at the circuit level, the transmitter generates the signals based on direct interactions between the spatial waves and the STC metasurface, and hence it does not need the RF components such as mixers. At the receiving end, the STC metasurface is used to capture the echo waves. More importantly, we develop a theory to implement the dechirp calculation by metasurface, which means that the dechirp process is realized on the RF level, avoiding the conventional RF hardware. The STC-metasurface-based RF dechirp processing transforms the broadband chirp signals into narrowband signals, significantly lowering the otherwise prohibitively high AD sampling rates. This advancement can simplify the down-conversion originally from the RF signals to baseband to be directly realized through the digital chains, making it easier and much less burdensome on the entire receiving chain. Compared to the conventional methods, this STC-metasurface-based configuration can enhance the signal processing efficiency, simplify the system architecture, and reduce the cost, complexity, and energy consumption. We remark that the roles of two STC metasurfaces can be interchanged due to their identical structures and reprogrammable nature, thus increasing the flexibility of the proposed radar system.

Several experiments are conducted to validate the detection and ranging capabilities of the STC-metasurface-based radar system. Furthermore, our radar demonstrates its capability for speed measurement. To the best of our knowledge, this is the first instance to use the information metasurfaces as the core components in constructing a radar system with the STC capabilities, information modulation capacities, beamforming features, and RF calculation functionalities. This new and simple STC-metasurface-based radar architecture is expected to offer new solutions for emerging fields such as integrated communication and sensing and smart radar.

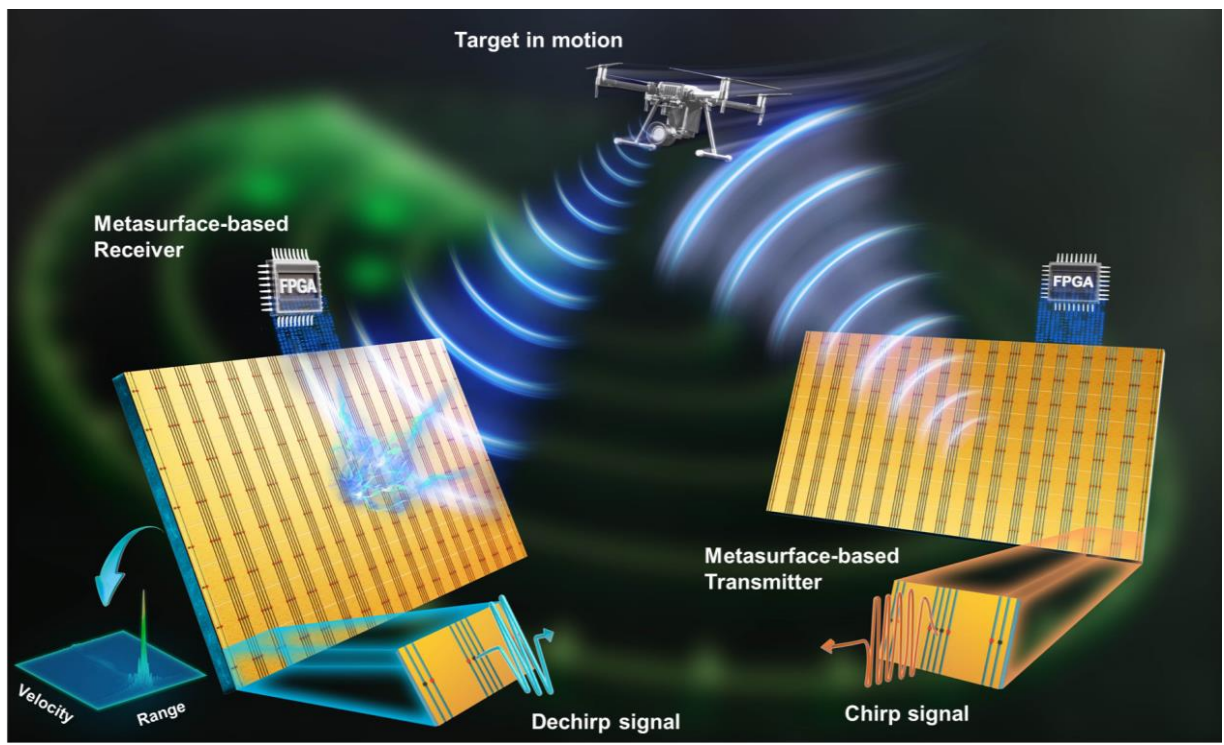

Fig. 1. Conceptual diagram to illustrate the STC-metasurface-based radar system. One STC metasurface is responsible for generating the RF chirp signals and transmitting them to targets, while the other STC metasurface receives the echo signals and performs the RF dechirp processing directly on the metasurface.

# Theory and results

**STC-metasurface-based radar architecture**

In the proposed radar system, two distributed STC information metasurfaces are employed as key components at the transmitting and receiving ends, as shown in Fig. 1. This configuration can effectively prevent the transmitting signals from entering the receivers when the information metasurfaces are used for the RF signal generation and RF dechirp processing in space. From the perspective of information metasurfaces, using the multi-metasurface control techniques can facilitate the synchronization of the transmitter and receiver in the presented radar. The two STC metasurfaces are designed with identical structures, including the unit cells and controlling circuits, to ensure the consistency of the chirp signals and the perfect matching filter operation.

The operation of the STC-metasurface-radar system is described as follows. The STC metasurface at the transmitting end directly converts the monochromatic plane waves into RF chirp signals and sends them out into space. When these signals impinge on targets, they are scattered as radar echoes. The echo signals are then received by the other STC metasurface, which features time-varying reflectivity and undergoes direct calculations to achieve the dechirp operation on the physical RF level. Subsequently, the data are reorganized into a two-dimensional (2D) matrix in fast- and slow-time. Finally, they are calculated by a 2D fast Fourier Transform (FFT) to generate the range and speed results of the targets.

The STC information metasurface consists of a lattice of unit cells, in which each unit is embedded with electrically tunable varactor diodes, as illustrated in Fig. 1. By controlling the reverse biasing voltages across these diodes, the reflection phase of the unit cell can be adjusted with a range greater than $2\pi$. Details on the cell structure and its full-phase control are introduced in Methods and Supplementary Note 1. When connected to a digital module and controlled by time-coding signals, the STC metasurface can tailor the EM waves in the frequency domain. In other words, it can produce multiple frequencies and manipulate their phases precisely, meeting the crucial prerequisites for chirp generation and dechirp behavior in this work.

According to the principle of STC metasurface, harmonic frequencies can be generated by altering the reflection phase of the unit cell over time[31]. To further introduce a phase shift at the $m^{\text{th}}$-order harmonic, a time shift $t_d$ is produced to the modulation waveform. Then the phase shift becomes $\varphi = -2\pi m f_m t_d$, where $f_m$ is the modulation frequency. To demonstrate the STC metasurface's abilities to generate harmonics and tune their phases, the scattering characteristics of the 1$^{\text{st}}$-order harmonic are studied as an example. Based on the array theory, a phase gradient should be formed across the metasurface to realize a beam deflection angle $\theta_{des}$ in the far-field region, and the harmonic phase difference between the neighboring meta-

columns should be

$$\Delta\varphi = \frac{2\pi f_o d}{c} \cdot \sin\theta_{des},$$ (1)

where $c$ denotes the light speed in free space, $f_o$ represents the operational frequency, and $d$ accounts for the distance between the adjacent meta-columns.

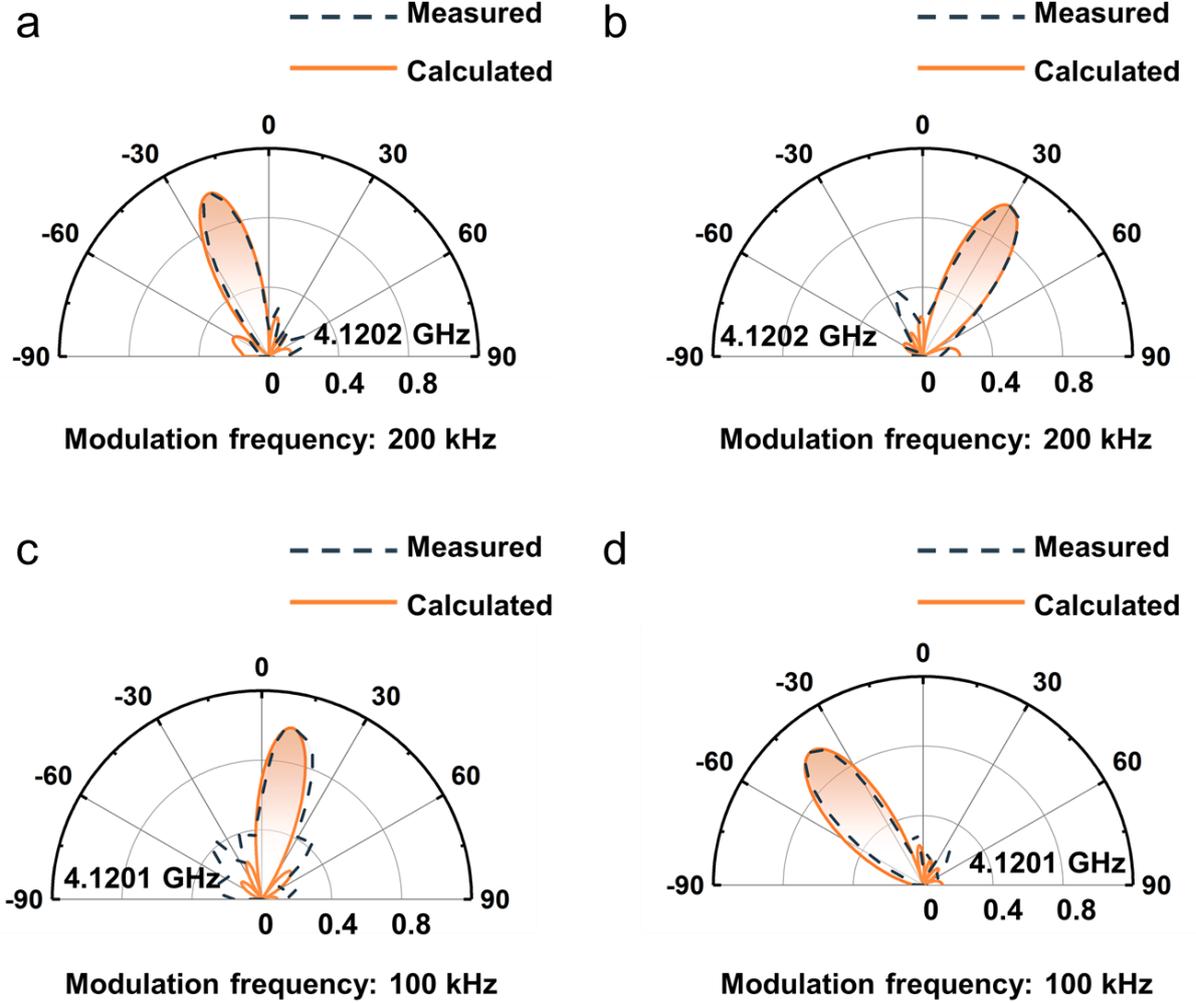

Fig. 2. The first-harmonic beamforming properties realized by the STC metasurface, in which the incident frequency is 4.12GHz. (a-b) The beam deflection angles of -20° (a) and 30° (b) when the modulation frequency is 200 kHz. (c-d) The beam deflection angles of 10° (a) and -40° (b) when the modulation frequency is 100 kHz.

The metasurface under test in this section has eight meta-columns, each spaced 24 mm apart, and the incident frequency is 4.12 GHz. Firstly, we set the modulation frequency of the controlling signals for the STC metasurface to 200 kHz. To realize two steering angles of -20° and 30°, respectively, space phase gradients are purposely designed by time-shifting the controlling waveform. Figs. 2a and 2b display the calculated and measured scattering beams at 4.1202 GHz, which agree well with each other. In the same manner, we also demonstrate the first-harmonic beams steered to angles 10° and -40° with good performance when the modulation frequency is 100 kHz, as shown in Figs. 2c and 2d. Please refer to Supplementary Note 2 for more details on the designs of harmonic space phase gradients in these examples.

The active beamforming capability of the STC information metasurface is similar to that of the conventional phased arrays but avoiding the expensive and complex transmitter and receiver (T/R) components. With this feature, the proposed metasurface-based radar can be further optimized to measure the spatial angles of targets with a high resolution. Unlike recent studies on the direction of arrival (DOA) estimations using metasurfaces[55,56], which primarily rely on passive detection, our approach actively emits the EM waves toward the target. This active emission enables precise control over the transmitting waveform, frequency, and timing, thereby enhancing the target detection capabilities across diverse environments. In the nest section, we will concentrate on the target detection and range/velocity measurements.

**The STC-metasurface-based receiver**

As illustrated in Fig. 1, the primary function of the STC metasurface situated in the receiving section is to directly execute the RF dechirp processing on the metasurface, and deliver the resulting multi-tone signals to post-stage circuits to acquire the target's range and velocity. When the scattered echoes from targets arrive at the receiving metasurface, the signals are reflected again, and the electric field $E_{r1}$ is written as

$$E_{r1} = \Gamma_1 \cdot E_{i1}, \tag{2}$$

where $\Gamma_1$ is the time-varying reflectivity of the STC metasurface that can be regulated by the controlling voltages. $E_{i1}$ describes the electric field of the signal impinging on the receiving metasurface, which is the echo from the target. In the field of radar engineering, it can be presented as[1,4]

$$E_{i1}(t,k) = \sum_{n=1}^{N} \sigma_n e^{j\left(2\pi f_c\left(t - \frac{(R_n - 2(k-1)v_n T_p)}{c}\right) + j\pi \frac{B}{T_p}\left(t - \frac{(R_n - 2(k-1)v_n T_p)}{c}\right)^2\right)}, \tag{3}$$

where $t$ represents the fast time, while $(k-1)T_p$ represents the slow time, in which $k$ is the $k^{th}$ chirp signal in one frame; $N$ represents the number of targets; $\sigma_n$ is proportional to the radar cross section (RCS) of the $n^{th}$ target, range attenuation, and so forth; $f_c$ is the carrier frequency; $R_n$ represents the transceiver range, that is, $R_n = R_t + R_r$, where $R_t$ and $R_r$ indicate the distances from the transmitter to the $n^{th}$ target and from the $n^{th}$ target to the receiver, respectively; $v_n$ denotes the radial velocity of the $n^{th}$ target; and $B$ and $T_p$ account for the bandwidth and duration of the FMCW signal, respectively. It should be mentioned that Eq. (3) is entirely distinct from the monochromatic plane-wave incidences considered in the previous research on metasurfaces, and it leads to the unique wave-signal-matter interaction on the STC metasurface for radar signal processing.

To realize the dechirp processing on the STC metasurface, we introduce a time-varying reflection phase to cancel out the phase term $\pi B/T_p$ in Eq. (3), while preserving the phase terms that contain the range/speed information. Specifically, the reflectivity $\Gamma_1$ is expressed as

$$\Gamma_1(t) = e^{-j\pi\frac{B}{T_p}t^2}. \tag{4}$$

From Eq. (4), it can be seen that the amplitude of $\Gamma_1$ is a constant, and the phase of $\Gamma_1$ is a quadratic function of time. More importantly, the reflection phase of $\Gamma_1$ should undergo $2\pi$ variation in each cycle, which raises an essential requirement for the metasurface to be used. With the mapping relationship between the biasing voltage and reflection phase of the STC metasurface, $\Gamma_1$ can be accurately regulated in real time. By substituting Eq. (3) and Eq. (4) into Eq. (2), $E_{r1}$ can be deduced as

$$E_{r1}(t,k) = \sum_{n=1}^{N} \sigma_n e^{j2\pi f_c t} e^{j\left[\frac{2\pi B R_n}{cT_p}t - \frac{2\pi(R_n - 2(k-1)v_n T_p)}{\lambda_c}\right]}. \tag{5}$$

Eq. (5) is the theoretical foundation of this work, which indicates that the dechirp processing is directly performed on the RF level. It also implies that the RF dechirp processing converts the broadband chirp signals into narrowband signals, substantially reducing the high sampling rates and thereby easing the demands on the entire receiving chain. Specifically, the baseband signal in Eq. (5) manifests a set of discrete frequency tones, where each frequency tone corresponds to a specific range based on the time delay of the reflected signal. When this baseband signal is processed using FFT, the resulting frequency spectrum represents the one-dimensional (1D) range profile, in which the amplitude at each frequency indicates the strength of the reflection from targets at that particular range. Hence, the FFT operation at this stage is commonly referred to as the range FFT.

After detecting the $n^{\text{th}}$ target and measuring its distance, we proceed to calculate its velocity. To achieve this goal, the baseband signals from a complete frame are collected. Specifically, Eq. (5) is adjusted as

$$E_{r1}(t) = \sum_{k=1}^{K} \sigma_n e^{j2\pi\frac{B R_n}{cT_p}t} e^{-j\frac{4\pi(k-1)v_n T_p}{\lambda_c}}. \tag{6}$$

Eq. (6) indicates that $K$ returns are obtained from the $n^{\text{th}}$ target. These $K$ signals share the same frequency but have different phases. The measured phase difference $\Delta\varphi = 4\pi v_n T_p/\lambda_c$ is the function of velocity of the $n^{\text{th}}$ target. Subsequently, a second FFT, referred to as the Doppler FFT, is applied to the $K$ phasors. This allows the velocity of the $n^{\text{th}}$ target to be calculated as $\Delta\varphi\lambda_c/4\pi T_p$.

In conclusion, the STC-metasurface-based receiver directly performs the dechirp signal processing on the physical platform of the metasurface, and the signals reradiated by the STC metasurface can be further processed using 2D FFT to determine the ranges and speeds. Please refer to Supplementary Note 3 for more details. The feasibility of this method will be validated in the experimental section.

## The STC-metasurface-based chirp signal transmitter

The ability to transmit the chirp signals into space is the key to an FMCW radar system. As shown in Fig. 1, the transmitting STC metasurface can generate the desired FMCW signals directly. Here, the monochromatic EM waves at the frequency of $f_c$ are normally incident to the metasurface, and the reflected electric field $E_{r2}$ is described as

$$E_{r2} = \Gamma_2 \cdot e^{j2\pi f_c t}, \tag{7}$$

in which $\Gamma_2$ represents the time-varying reflection coefficient of the STC metasurface. For the chirp generation, $\Gamma_2$ should be regulated as

$$\Gamma_2(t) = e^{j\pi \frac{B}{T_p} t^2}, \tag{8}$$

in which $B$ and $T_p$ represent the bandwidth and duration of the FMCW signal, respectively. It can be seen from Eq. (8) that the reflection phase of $\Gamma_2$ experiences a 360-degree variance in a single period, which poses a fundamental criterion for the effective functionality of the STC metasurface.

To validate the ability of the STC metasurface to generate the chirp signals, a series of experiments are conducted. The bandwidth $B$ and duration $T_p$ of the signals are $100\,\text{kHz}$ and $100\,\mu\text{s}$, respectively. The experimental setup includes an STC metasurface controlled by a digital module, which converts the incident monochromatic signals into the chirp ones, and a universal software radio peripheral (USRP-2974) that down-converts the chirp signals to the baseband. More details can be found in Methods.

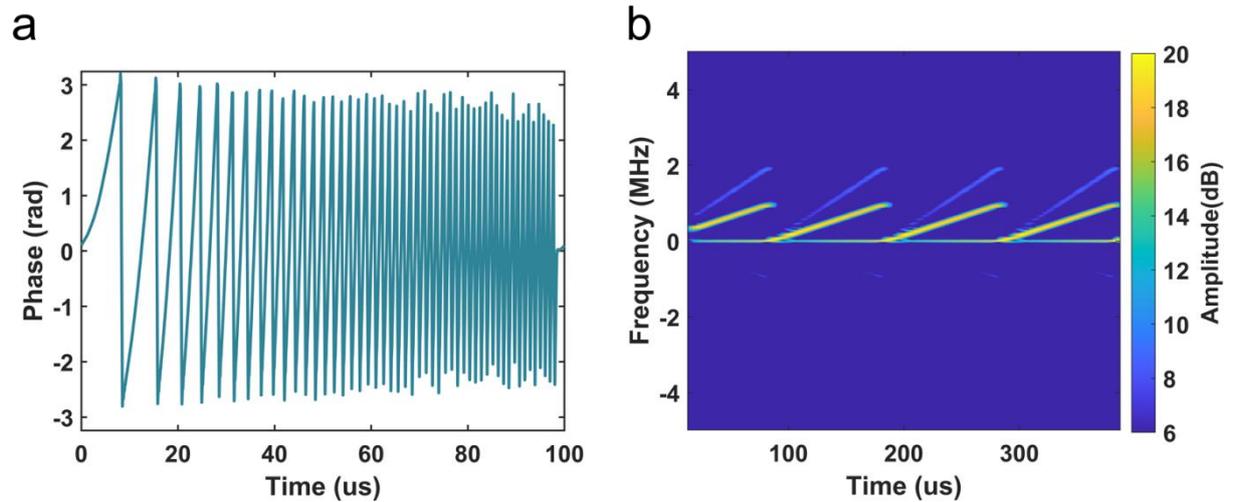

Fig. 3. The baseband chirp signal generated by the STC metasurface. (a) The phase of the generated chirp signal. (b) Segment of the time-frequency curve of the generated baseband signal.

The baseband signals are shown in Fig. 3a. In addition, Fig. 3b illustrates a segment of the chirp signals transmitted by the STC metasurface in the form of time-frequency curves. Figs. 3a and 3b indicate that the chirp signals are successfully produced by the metasurface. Besides the chirp signals, some undesired frequency components, such as the fundamental and high-order harmonics, are also reflected by the STC metasurface, as seen in Fig. 3b. The primary cause of the fundamental harmonic is the spatial multi-path effect; the high-order harmonics such as the $1^{st}$- and 2nd-order harmonics are mainly the result of fluctuations in the reflection amplitude of the metasurface and the nonlinear characteristics of the controlling components. Nevertheless, due to the identical structure and performance of the receiving and transmitting metasurfaces, the undesired frequency components generated by them are the same. The chirp signals produced by one metasurface can be effectively processed by the other as a matched filter while preserving the ranging and velocity measurement capabilities of the new system. It is also highlighted that the reprogrammable feature of the metasurface allows for dynamic regulations of the chirp signal parameters, including bandwidth, making it potentially useful for modern radar anti-jamming techniques.

**Experimental validation for the STC-metasurface-based radar system**

We conduct a series of experiments to validate the performance of the STC-metasurface-based radar system. The picture of the radar system is presented in Fig. 4. The transmitting section is shown on the left side. A single-tone signal at 4.25 GHz with an instantaneous bandwidth of 1 GHz is generated by a vector signal transceiver (VST, PXIe-5841), which is radiated by a horn antenna (ANT 1) to illuminate the transmitting metasurface. The chirp signal directly generated by the metasurface is transmitted into space. The receiving section is shown on the right side. The radar echoes experience the dechirp processing on the receiving metasurface, and the processed signals are then re-scattered and captured by a horn antenna (ANT 4). ANT 4 is connected to VST, where the signals are converted into baseband signals. Finally, the range and/or speed information of the targets are extracted. The controlling signals for the metasurfaces are provided by a 16-bit waveform generator (WG, PXIe-5433). The VST and WG modules are installed in a multi-slot chassis (PXIe-1092) along with an embedded controller (PXIe-8881), ensuring that the entire system is operated on the same oscillator and maintains coherence.

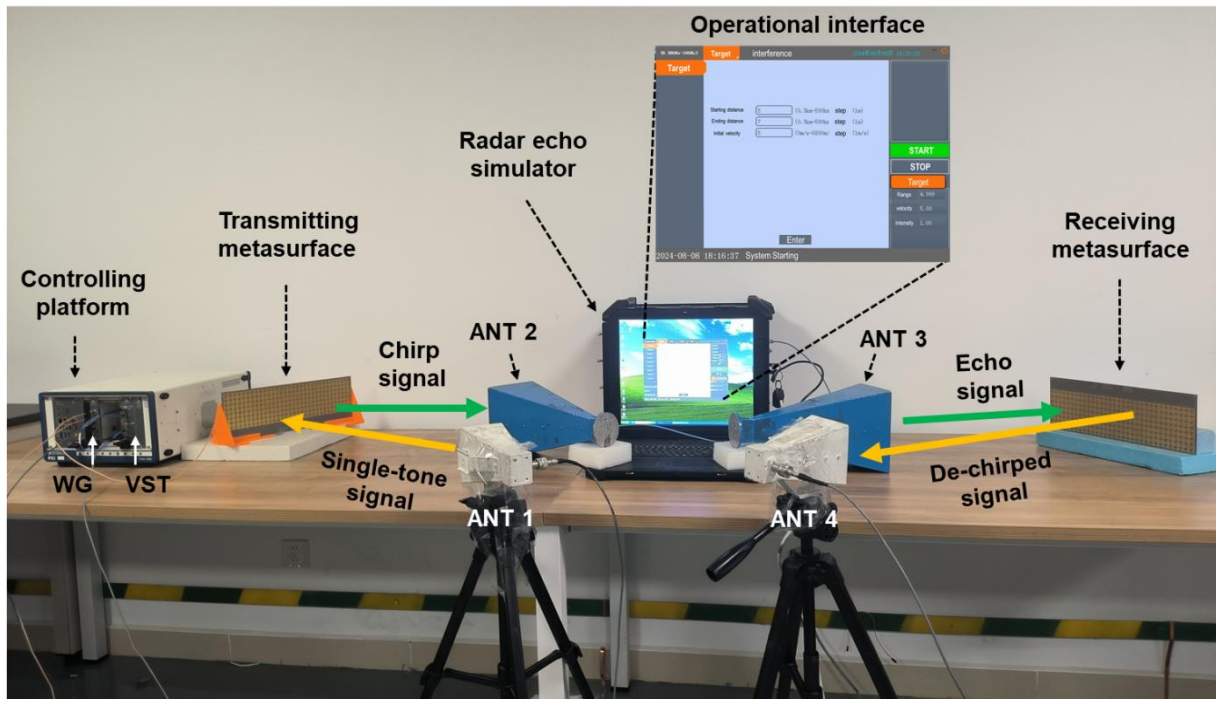

Fig. 4. The STC-metasurface-based radar system in an indoor scenario.

In this indoor radar testing environment configuration, a radar echo simulator (details are provided in Supplementary Note 4) receives the signals from the transmitting metasurface via a horn antenna (ANT 2) and re-radiates the echoes encoded with specific motion information of the targets through another horn antenna (ANT 3). In the experiments, the transmitting metasurface emits chirp signals with a bandwidth of 1 MHz, resulting in a range resolution of 0.15 km. Besides that, the pulse width is 100 ms, and each coherent processing interval (CPI) consists of 128 chirps. In modern radar systems, it is essential for the transmitting and receiving RF signals to be coherent. Therefore, the low-frequency controlling signals for the two STC metasurfaces are synchronized with a common local oscillator. At the same time, a conventional radar system is used for comparison, which consists of two antennas, RF chains (including components such as a low-noise amplifier and mixer), and digital processing units. It transmits 128 chirps within a single CPI, each with a 1 MHz bandwidth and a 100 ms pulse width. Theoretically, this configuration yields the same range and velocity resolutions as the STC-metasurface-based radar.

The first experiment is conducted to illustrate the single-target range measurement using the metasurface-based radar. In the radar echo simulator, a stationary target is set at a distance of 1 km for testing. The result from the proposed radar (in green) is depicted in Fig. 5a, showing the measured distance of 1.09 km, with a measuring error of 0.09 km that falls below the distance resolution of 0.15 km. The measured result from the conventional radar (in green) is 1.09 km too, as shown in Fig. 5b, which is quite close to the result of the proposed radar.

Then, to validate the joint range-velocity measurement capability of the presented radar system, we set a target with a range of 1 km and a velocity of 10 m/s. In Figs. 5(c-d), the measured results of the proposed radar and the conventional radar are compared. As described in Fig. 5c, the target's range and velocity are set to 1 km and 10 m/s, respectively, and the measured values from the proposed radar are 1.02 km and 10.04 m/s. The resulting errors in range and velocity, 0.02 km and 0.04 m/s, are both below their respective resolutions, which demonstrates the capability of the STC-metasurface-based radar for accurate measurement of the range and velocity for a single target. As for the conventional radar, the measured results are the same as those from the proposed radar, as shown in Fig. 5d.

In practical applications, it is common to have multiple targets that need to be detected simultaneously. Here, we consider two dual-target cases. In Case I shown in Fig. 5e and 5f, the two targets are set as (3 km, 0 m/s), (1 km, 0 m/s); while in Case II shown in Fig. 5g and 5h, the two targets are set as (3 km, 0 m/s), (1 km, 10 m/s). Figs. 5e and 5g depict the results measured by the metasurface-based radar, and Figs. 5f and 5h show the results obtained by the conventional radar. It can be found that the presented radar has successfully detected all preset targets. Specifically, the measured results in Case I are (3.01 km, -0.03 m/s) and (1.04 km, 0.03 m/s), with the errors below 0.04 km and 0.03 m/s, respectively; and the results in Case II are (3.01 km, 0.03 m/s) and (1.06 km, 9.97 m/s), with the errors below 0.06 km and 0.03 m/s, respectively. These results are almost the same as those from the conventional radar. These experiments have demonstrated the satisfactory capabilities of the proposed radar system for target detection and measurement. To enhance the transmission signal bandwidth, a more advanced metasurface prototype can be designed to accommodate significantly higher modulation speeds. This will be studied in our future work.

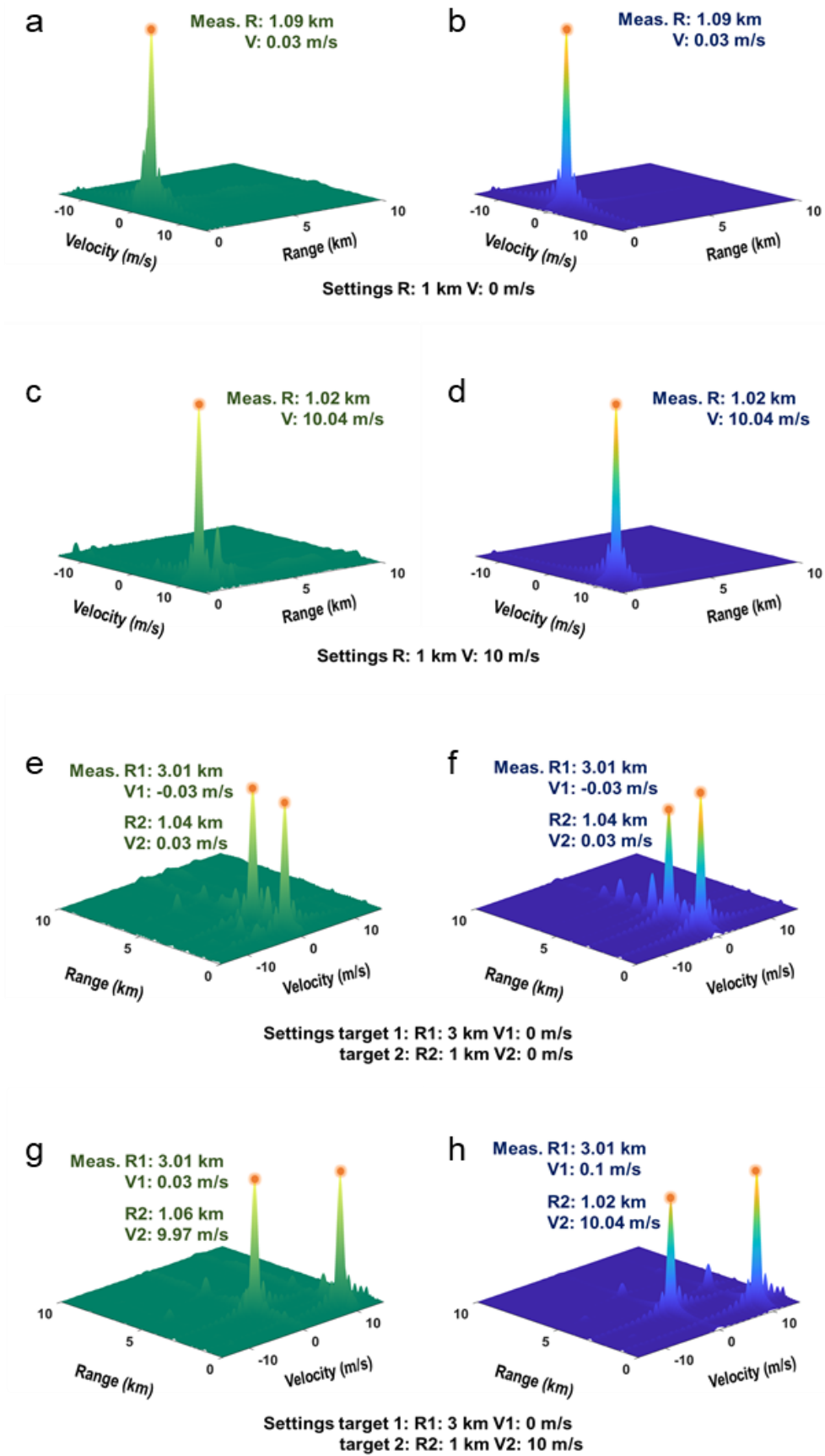

Fig. 5. The ranges (R) and velocities (V) measured by the STC-metasurface-based radar (displayed in green) and the conventional radar (displayed in blue), respectively. (a), (c), (e), and (g) depict the results of the metasurface-based radar, while (b), (d), (f), and (h) showcase the results of the conventional radar. The measurement settings: (a)(b). (1 km, 0 m/s); (c)(d). (1 km, 10 m/s); (e)(f). (3 km, 0 m/s), (1 km, 0 m/s); (g)(h). (3 km, 0 m/s), (1 km, 10 m/s). The measured results: (a). (1.09 km, 0.03 m/s); (b). (1.09 km, 0.03 m/s); (c). (1.02 km, 10.04 m/s); (d). (1.02 km, 10.04 m/s); (e). (3.01 km, -0.03 m/s), (1.04 km, 0.03 m/s); (f). (3.01 km, -0.03 m/s), (1.04 km, 0.03 m/s); (g). (3.01 km, 0.03 m/s), (1.06 km, 9.97 m/s); (h). (3.01 km, 0.1 m/s), (1.02 km, 10.04 m/s)

## Discussion

We propose a novel metasurface-based radar system that leverages the STC metasurfaces in both transmitter and receiver. The STC metasurface in the transmitter generates the FMCW signals from monochromatic EM wave incidences and sends them into space, eliminating the need for the conventional VCO or DDS technology. Meanwhile, the STC metasurface in the receiver captures the echo signal and performs the RF dechirp calculations without using the traditional RF circuits or algorithms. It offers a metasurface-based approach for processing the radar signals at the RF front end, which holds several advantages, including improved signal quality, increased processing speed, and reduced power consumption. Several experiments are carried out for the detection and measurement of single and dual targets, and their distances and velocities are successfully acquired with controllable range and velocity resolutions. The results are on par with those from the conventional radar systems, yet they are achieved with significantly simplified hardware and reduced costs. This methodology has demonstrated promising potential for future applications in radar, integrated communication, and sensing technologies.

## Methods

### Details on the metasurface and measurement

The metasurface employed in this study operates at approximately 4.25 GHz and comprises an 8×16 array of meta-atoms. The meta-atom features a three-layered structure, as depicted in Fig. 6a, with the top layer incorporating four chip capacitors and four varactor diodes that bridge adjacent metallic strips. By varying the bias voltage applied to the varactor diodes from 0 to 13 V, the reflection phase was measured. The relationship between the biasing voltage and the reflection phase is illustrated in Fig. 6b. It can be seen that the reflection phase range exceeds $2\pi$ as the biasing voltage increases, facilitating the generation of arbitrary signals based on the measured mapping relationship. It is apparent that the meta-atom's full-phase control capability meets the requirements for chirp signal generation and dechirp calculation, as indicated in Eq. (8) and Eq. (4). Simulation details of the meta-atom are provided in Supplementary Note 1.

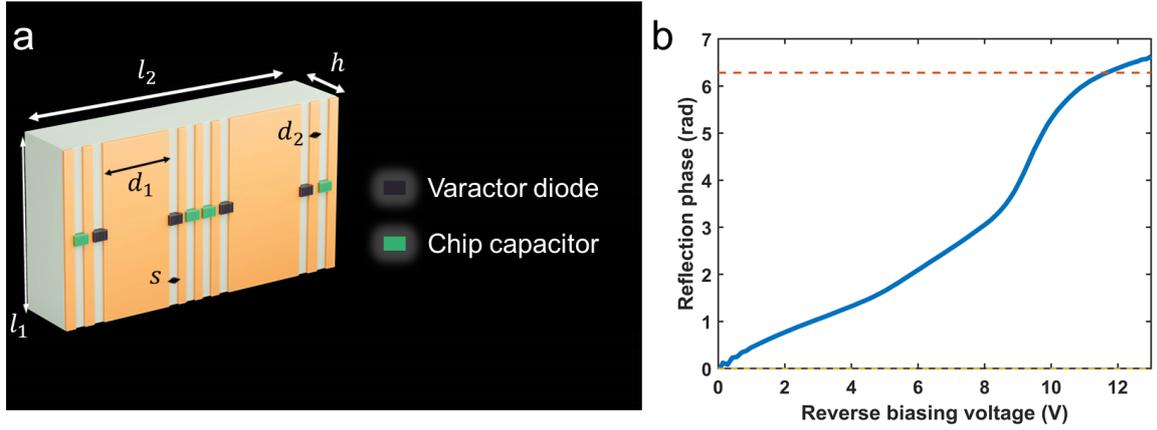

Fig. 6. The meta-atom and measured features. (a) Outline of the meta-atom backed by a metallic plate, in which $l_1 = 12$ mm; $l_2 = 24$ mm; $h = 5$ mm; $d_1 = 5.6$ mm; $d_2 = 1.2$ mm; and $s = 0.7$ mm. (b) The measured reflection phases of the metasurface by increasing the reversely biasing voltage.

**Experimental setup to validate the chirp signal generation**

The experiment was conducted in an indoor scenario. The setup can be categorized into two sections: the transmitting end and the receiving end. At the transmitting end, the STC metasurface was controlled by an NI chassis consisting of a high-speed I/O bus controller, an FPGA module, a digital-analog conversion (I/O) module, a DC power supply module, etc. A horn antenna is linked to a microwave signal generator (Keysight E8267D) to send monochromatic excitation signals at 4.25 GHz to the metasurface. At the receiving end, another horn antenna is connected to a universal software radio peripheral (USRP-2974) to capture the chirp signals reflected by the STC metasurface and inject them into USRP, which down-converts the chirp signals to the baseband.

**Experimental setup for the reflection phase measurement**

The experiment was conducted in a microwave anechoic chamber, with the metasurface being controlled by the NI custom chassis (PXIe-1092) comprising a high-speed I/O bus controller, an FPGA module (PXIe-7976R), a timing (PXIe-6674T), an I/O module (PXIe-5783), etc. On the other side, a vector network analyzer was used to receive the reflected signals via a horn antenna and measure the corresponding reflection coefficient $S_{11}$. As the biasing voltage increases from 0 V to 13 V, the phases of $S_{11}$ are recorded. The mapping relationship between them is the reflection phase-voltage mapping relationship.

**Data availability**

The authors declare that all relevant data are available in the paper and its Supplementary Information Files, or from the corresponding author on request.

## Code Availability

The custom computer codes utilized during the current study are available from the corresponding authors on request.

## Acknowledgments

This work is supported by the National Key Research and Development Program of China (2023YFB3811502, 2018YFA0701904), the National Science Foundation (NSFC) for Distinguished Young Scholars of China (62225108), the Program of Song Shan Laboratory (Included in the management of Major Science and Technology Program of Henan Province) (221100211300-02, 221100211300-03), the National Natural Science Foundation of China (62288101, 62201139, U22A2001, 62471134), the Jiangsu Province Frontier Leading Technology Basic Research Project (BK20212002), the Natural Science Foundation of Jiangsu Province (BK20221209), the Fundamental Research Funds for the Central Universities (2242022k60003, 2242023K5002), and the Southeast University - China Mobile Research Institute Joint Innovation Center (R202111101112JZC02).

# Supplementary Information

# Simplified radar architecture based on information metasurface

Si Ran Wang[1], Zhan Ye Chen[1,2,3,*], Shao Nan Chen[1,2,3], Jun Yan Dai[1,2,3], Jun Wei Zhang[1,2,3], Zhen Jie Qi[1,2,3], Li Jie Wu[1,2,3], Meng Ke Sun[1,2,3], Qun Yan Zhou[1,2,3], Hui Dong Li[1,2,3], Zhang Jie Luo[1,2,3,*], Qiang Cheng[1,2,3,*], Tie Jun Cui[1,2,3,4,*]

1. State Key Laboratory of Millimeter Waves, Southeast University, Nanjing 210096, China
2. Institute of Electromagnetic Space, Southeast University, Nanjing 210096, China
3. Frontiers Science Center for Mobile Information Communication and Security, Southeast University, Nanjing 210096, China
4. Suzhou Laboratory, Suzhou 215164, China

These authors contributed equally: *Si Ran Wang, Zhan Ye Chen, Shao Nan Chen*

E-mail: chenzhanye@seu.edu.cn; zjluogood@seu.edu.cn; qiangcheng@seu.edu.cn; tjcui@seu.edu.cn

**Supplementary Note 1: Meta-atom design and simulation performances**

As illustrated in Fig. S1a, the meta-atom consists of a three-layer structure, with a middle F4B substrate ($\varepsilon_r$=3.0  tan $\sigma$=0.0015) sandwiched between the top and bottom copper layers. The top layer integrates four chip capacitors (0.1 pF, marked in dark blue) and four varactor diodes (SMV-2019, Skyworks, Inc., marked in red) that bridge adjacent metallic strips. By applying various reverse biasing voltages to the varactor diodes, the reflection phase of the meta-atom can be continuously tuned. Additionally, 16 metallic vias connect the top and bottom layers to extend the reflection phase range. The meta-atom's overall dimensions are 5 mm in thickness, 24 mm in length, and 12 mm in width.

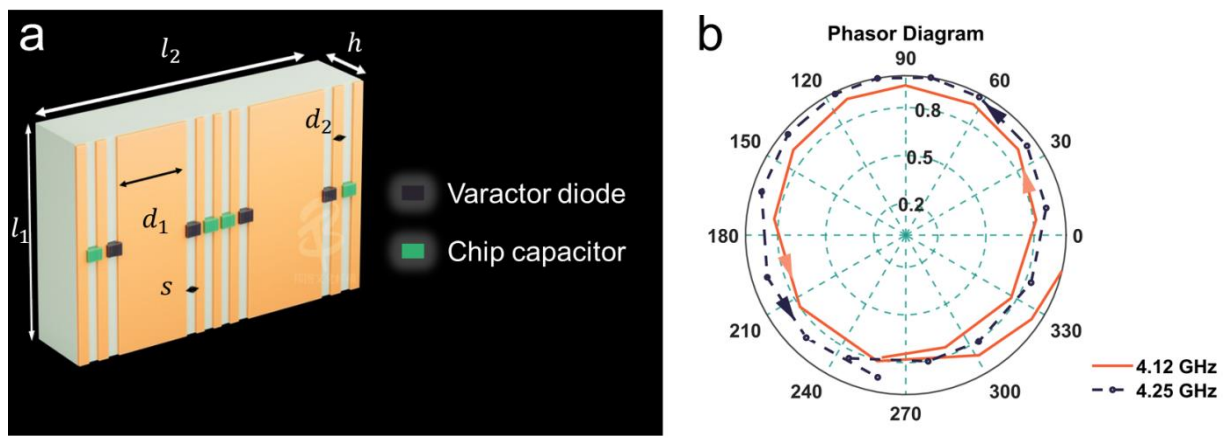

Fig. S1. Outlines of the designed meta-atom and its simulated reflection phases and amplitudes under various reverse biasing voltages. (a) The outlines of the meta-atom. $l_1 = 12$ mm; $l_2 = 24$ mm; $h = 5$ mm; $d_1 = 5.6$ mm; $d_2 = 1.2$ mm; $s = 0.7$ mm. (b) The reflectivity phasor diagram of the meta-atom under a linearly increasing bias voltage (counterclockwise) at frequencies of 4.12 GHz and 4.25 GHz, respectively

Full-wave simulations for the meta-atom are conducted using the commercial software, CST Microwave Studio 2016. The reflectivity phasor diagram of the meta-atom under a linearly increasing bias voltage (counterclockwise) at 4.12 GHz and 4.25 GHz, as shown in Fig. S1b, reveals that the reflection phase range exceeds $2\pi$.

**Supplementary Note 2: Beamforming at the harmonic frequencies of 4.1201 and 4.1202 GHz**

The harmonic scattering patterns are measured in a microwave anechoic chamber. A horn antenna operating in the C-band is the excitation source, connected to a microwave signal generator (Keysight E8267D) that produces a monochromatic signal at 4.12 GHz. The metasurface and the feeding antenna are mounted on a turntable capable of automatic 360° rotation in the horizontal plane. On the receiving end, another horn antenna captures the scattered signals, which are recorded using a vector network analyzer.

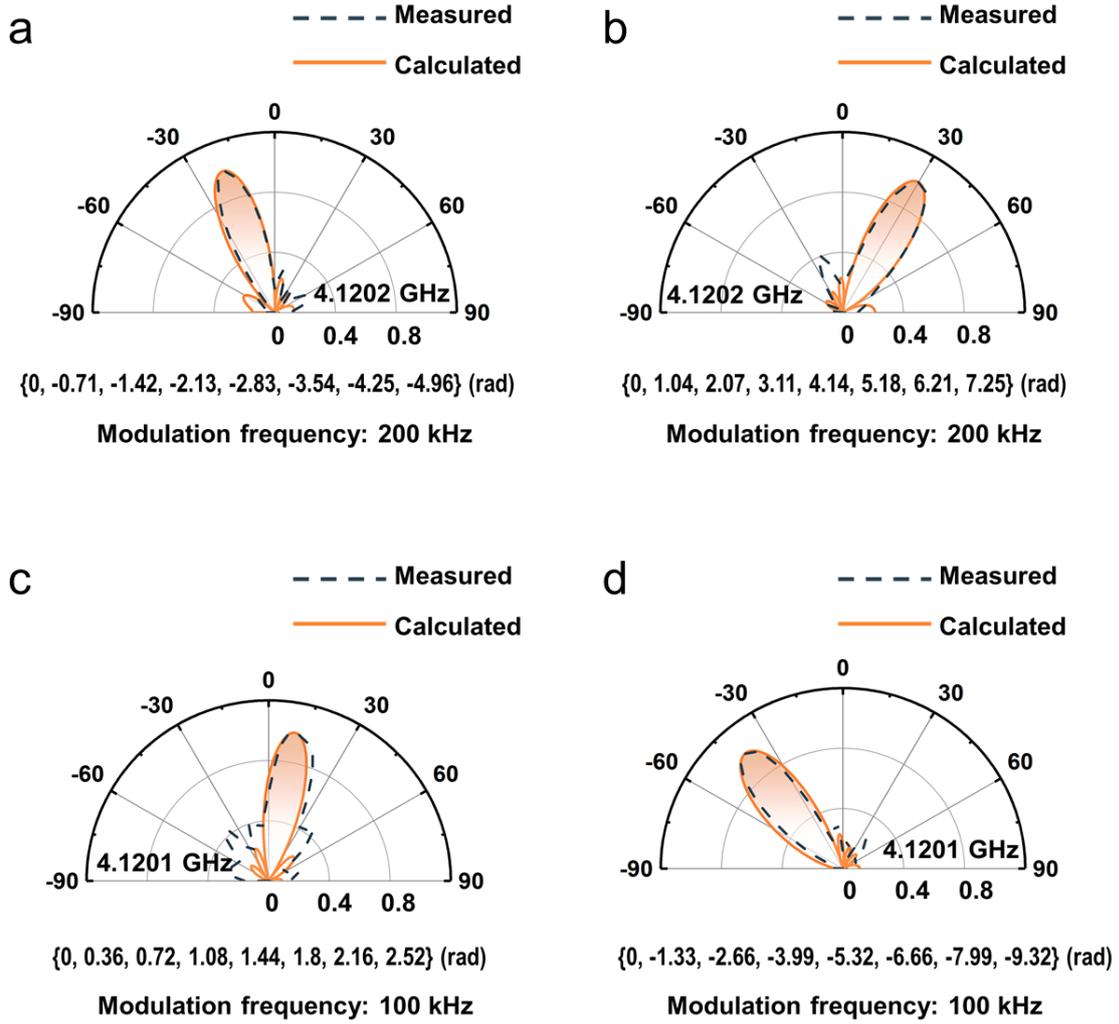

Fig. S2. Beamforming achieved by the STC metasurface at the 1st-order harmonic. (a-b) For a modulation frequency of 200 kHz: (a) The measured main lobe is at -20°; (b) The measured main lobe is at 30°. (c-d) For a modulation frequency of 100 kHz: (c) The measured main lobe is at 10°; (d) The measured main lobe is at -40°. The harmonic phase distributions that account for the scattering phenomena are presented along with the patterns

We use an STC metasurface prototype with eight meta-columns and interelement spacing of 24 mm to implement the beamforming at the 1st-order harmonic. It is controlled by the temporal modulation under the incident frequency of 4.12 GHz. Firstly, the modulation frequency is defined as 200 kHz, so the operational frequency is 4.1202 GHz. The desired angles of the main beams are -20° and 30°, respectively. According to Eq. (1) in the main text, the phase distributions should be {0, -0.71, -1.42, -2.13, -2.83, -3.54, -4.25, -4.96} (rad) and {0, 1.04, 2.07, 3.11, 4.14, 5.18, 6.21, 7.25} (rad), respectively, as shown in the figures. The corresponding phase gradients can be realized by introducing proper time shifts to the coding

waveforms. Specifically, the coding waveform used can be given by

$$C(t) = e^{jp \times \mathbf{mod}(t,T)}, \tag{S1}$$

in which $\mathbf{mod}(t,T)$ denotes the remainder of $t/T$ and $p$ is the phase slope. According to ref. S1, when $pT = 2\pi$, the incident frequency will be perfectly shifted to the 1st-order harmonic. As shown in Figs. S2a and S2b, the scattering beams at 4.1202 GHz are steered to -20° and 30°, respectively.

Subsequently, the modulation frequency is defined as 100 kHz, so the operational frequency is 4.1201 GHz. The desired angles of the main beams are 10° and -40°, respectively. Accordingly, the phase distributions are {0, 0.36, 0.72, 1.08, 1.44, 1.8, 2.16, 2.52} (rad) and {0, -1.33, -2.66, -3.99, -5.32, -6.66, -7.99, -9.32} (rad), respectively, as shown in the figures. Figs. S2c and S2d illustrate the scattering beams at 4.1201 GHz being steered to angles of 10° and -40°. It is evident that the measured results (dotted lines) agree with the simulations (solid lines), as illustrated in Fig. S2.

## Supplementary Note 3: Fundamental principle of frequency-modulated continuous wave (FMCW) radar

The FMCW radar's[S2] transmitting signal, known as the chirp signal, can be given by

$$s_t(t) = rect\left(\frac{t}{T_p}\right) e^{j2\pi\left(f_c t + \frac{B}{2T_p}t^2\right)}, \tag{S2}$$

in which $f_c$ represents the carrier frequency, $T_p$ and $B$ denote the duration and bandwidth of the chirp, respectively, and $rect(u)$ is described as

$$rect(u) = \begin{cases} 1, |u| \leq \frac{1}{2} \\ 0, |u| > \frac{1}{2} \end{cases}. \tag{S3}$$

Then, a target echo can be given by

$$s_r(t,k) = \sigma rect\left(\frac{t-\tau}{T_p}\right) e^{j2\pi\left(f_c(t-\tau) + \frac{B}{2T_p}(t-\tau)^2\right)}, \tag{S4}$$

where $\sigma$ reflects the scattering characteristic of the $n$th target, while $\tau$ is the round-trip time of the target, that is

$$\tau = \frac{2R_0 + 2v_0(k-1)T_c}{c}, \tag{S5}$$

in which $R_0$ represents the distance from the radar to the target. $v_0$ accounts for the velocity of the target. $k$ denotes the $k$th chirp signal in one coherent processing interval (CPI). $T_c$ represents the interval between two adjacent chirps. $(k-1)T_c$ can be called the slow time. Accordingly, $t$ is the fast time. Assume $T_c = T_p$. $c$ is the light speed in free space.

Suppose that the systemic reference range is $R_f = 0$. Thus, the reference signal can be

written as

$$s_f(t) = rect\left(\frac{t}{T_p}\right) e^{j2\pi\left(f_c t + \frac{B}{2T_p} t^2\right)}. \tag{S6}$$

Then, the dechirp processing is performed at the receiving end, which can be expressed mathematically as follows

$$s_{If}(t, k) = s_r(t, k) \cdot s_f^*(t), \tag{S7}$$

in which $*$ means a conjugate operation. $s_{If}$ can be expanded as

$$s_{If}(t, k) = A e^{j2\pi\left(\frac{2BR_0}{cT_p} t + \frac{2(R_0 + v_0(k-1)T_c)}{\lambda_c}\right)}, \tag{S8}$$

in which $A$ represents the signal amplitude after the dechirp processing. $\lambda_c = c/f_c$. For the sake of simplicity, we have assumed that the residual video phase has been disregarded. From Eq. (S8), $s_{If}$ is a single-tone signal with the frequency of $2BR_0/cT_p$ and the phase of $2(R_0 + v_0(k-1)T_c)/\lambda_c$. According to Ref. S1, $R_0$ and $v_0$ of the target can be acquired by performing a two-dimensional Fast Fourier Transform with respect to fast and slow time.

## Supplementary Note 4: Radar echo simulator

The user interface of the radar echo simulator is shown in Fig. S3. This simulator allows for the integration of user-defined spatiotemporal data for multiple targets into the transmitted signals, which are then emitted as echo signals into the surrounding environment.

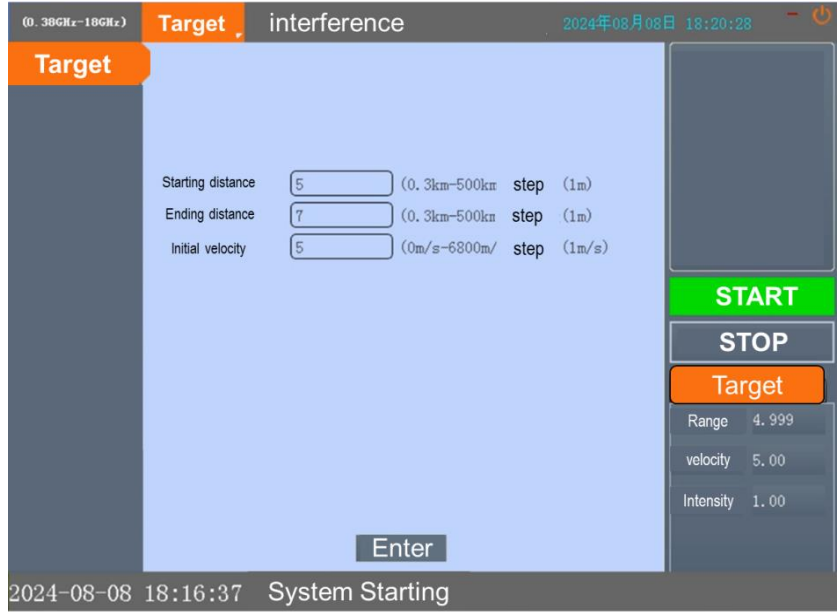

Fig. S3. The user interface of the radar echo simulator

To validate the reliability of the radar echo simulator, we employ a conventional FMCW radar with a chirp signal bandwidth of 1 MHz to receive the echoes generated by the simulator.

Three scenarios are tested using the radar echo simulator: (1) a single target at 1 km with a velocity of 0 m/s; (2) two targets at 1 km and 2 km, both with velocities of 0 m/s; (3) a single moving target at 1 km with a velocity of 10 m/s. The measured results are presented in Fig. S4. The radar successfully detects and measures the targets within the range resolution generated by the radar echo simulator, demonstrating that the simulator is a suitable tool for testing the performance of the metasurface-based radar system.

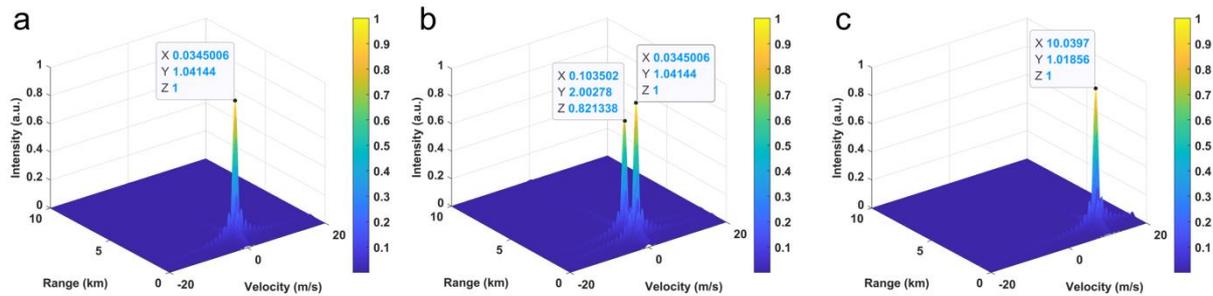

Fig. S4. The measured results of targets defined by the radar echo simulator. (a) Setting: (1 km, 0 m/s); measured: (1.02 km, 0.03 m/s). (b) Setting: (1 km, 0 m/s), (2 km, 0 m/s), measured: (1.04 km, 0.03 m/s), (2 km, 0.1 m/s). (c) Setting: (1 km, 10 m/s), measured: (1.01 km, 10.04 m/s)